\documentclass[a4paper,conference]{ieeeconf}

\IEEEoverridecommandlockouts     % (optional) for funding acknowledgments
\makeatletter
\newdimen\TargetLeft    \TargetLeft=19mm
\newdimen\TargetRight   \TargetRight=19mm
\newdimen\TargetTop     \TargetTop=25mm
\newdimen\TargetBottom  \TargetBottom=34mm

\AtBeginDocument{%
  \setlength\hoffset{0pt}%
  \setlength\voffset{0pt}%
  \setlength\headheight{0pt}%
  \setlength\headsep{0pt}%
  \setlength\footskip{12pt}%
  \setlength\oddsidemargin{\dimexpr \TargetLeft - 1in\relax}%
  \setlength\evensidemargin{\oddsidemargin}%
  \setlength\topmargin{\dimexpr \TargetTop - 1in\relax}%
  \setlength\textwidth{\dimexpr \paperwidth - \TargetLeft - \TargetRight\relax}%
  \setlength\textheight{\dimexpr \paperheight - \TargetTop - \TargetBottom\relax}%
}
\makeatother
\usepackage{graphicx}            % for images
\usepackage{amsmath,amssymb}     % math packages
\usepackage{url}                 % for clickable URLs

\usepackage{graphicx}      % include this line if your document contains figures
\usepackage{amsmath}      % For align, bmatrix, etc.
\usepackage{amssymb}      % For math symbols (\mathbb, \top, etc.)
\usepackage{bm}           % For bold math (\bm{})
\usepackage{mathtools} 
\usepackage{subcaption}

\usepackage{float} 

\usepackage{multirow}
\usepackage{comment}
\newcommand{\mysubsubsection}[1]{%
  \vspace{0.7em}\noindent\textbf{#1}\\[0.3em]\noindent
}
\usepackage{tikz}
\usetikzlibrary{
  arrows.meta,       % for '>=Stealth'
  positioning,       % for 'right=of' and 'below=of'
  shapes,            % for rectangles, rounded corners, etc.
  calc,              % for coordinate calculations
  backgrounds,       % optional, useful for layering
  fit                % for grouping nodes, if needed later
}
\usepackage{xcolor}
\definecolor{light}{RGB}{230,230,250} % soft lavender color (you can change it)

\begin{document}

\title{LPV Control for Dynamic Power Capping in High-Performance Computing under Mixed Workloads}

\author{
    \authorblockN{Mohamed Abdeldjalil Maziz}
    \authorblockA{
        Univ. Grenoble Alpes, Inria , GIPSA-lab\\
        F-38000 Grenoble, France\\
        \texttt{mohamed-abdeldjalil.maziz@inria.fr}
    }
    \and
    \authorblockN{Kouds Halitim}
    \authorblockA{
        Univ. Grenoble Alpes, Inria, CNRS, Grenoble INP, LIG\\
        F-38000 Grenoble, France\\
        \texttt{Kouds.Halitim@inria.fr}
    }
    \and
    \authorblockN{Bogdan Robu}
    \authorblockA{
        Univ. Grenoble Alpes, CNRS, Grenoble INP, GIPSA-lab\\
        F-38000 Grenoble, France\\
        \texttt{Bogdan.Robu@univ-grenoble-alpes.fr}
    }
    \and
    \authorblockN{Sophie Cerf}
    \authorblockA{
        Univ. Lille, Inria, CNRS, UMR 9189 CRIStAL\\
        F-59000 Lille, France\\
        \texttt{Sophie.Cerf@inria.fr}
    }
}

\maketitle
\begin{abstract} Balancing energy consumption and performance remains a critical challenge in High Performance Computing (HPC) systems. While static power capping mechanisms such as Intel’s Running Average Power Limit (RAPL) offer basic control capabilities, they lack the flexibility to adapt to dynamically varying workloads. This work studies dynamic power regulation for mixed workload scenarios. We investigate two feedback strategies: a gain scheduled proportional–integral (PI) controller and a polytopic linear parameter-varying (LPV) controller synthesized via $\mathcal{H}_\infty$ control, both scheduled by a workload indicator that changes between memory and compute phase. We evaluate tracking performance, phase switching, and robustness under practical power cap constraints. While both controllers respect power limits, the LPV design achieves lower tracking error, lower control variance, and smoother transients during phase changes than gain scheduled PI.
\end{abstract}


%\end{frontmatter}
%===============================================================================

\section{Introduction}
High-Performance Computing (HPC) has become an indispensable tool in advancing science and engineering from climate modeling to computational chemistry and beyond~\cite{ravanthi2014}. It enables large-scale simulations, data analysis, and machine learning tasks that exceed the capacity of conventional computing systems. Modern HPC systems are built upon massively parallel distributed architectures composed of millions of CPU and GPU cores~\cite{article1}, collectively delivering performance in the petaflop/s to exaflop/s  range (e.g., El-Capitan reaches $1.742*10^{18}$ floating point operations per second on the HPL benchmark)~\cite{top5002024}. This extreme capability comes at the cost of substantial energy consumption as 196 of the TOP500 supercomputers collectively consume more than 3 billion kilowatt-hours of electricity annually, comparable to the energy consumption of 800 000 households presenting both economic and environmental concerns. As a result, improving energy efficiency has become a central challenge for HPC~\cite{Silva2024decarbonization}.

To manage and reduce the energy consumption of HPC systems, several power management techniques have been proposed, including Dynamic Voltage and Frequency Scaling (DVFS) and Dynamic Power Management (DPM) %\BR{no need to put them is bold as we do not use them, right?}  
which adjust the power usage of system components based on workload activity~\cite{KUMBHARE2020102686,845896}. Although effective in principle, these techniques present notable limitations in HPC environments as they cannot explicitly enforce strict power limits at the hardware level, and their operation often depends on deep integration with the operating system or job scheduler. This dependency reduces portability and flexibility across architectures, making these approaches less practical~\cite{SAFARI2018311}.

In contrast, power capping has emerged as a practical and enforceable mechanism for power regulation. It imposes hardware-level upper bounds on the power usage of system components such as CPUs and memory, thereby maintaining thermal and operational constraints while limiting energy costs~\cite{Ramesh2019PowerCapping}. Among these mechanisms, Intel’s \textbf{RAPL} has gained wide adoption in HPC research and practice,~\cite{6702684,ostapenco:hal-04742418, 8853776} among others. RAPL allows users to configure a power limit and a time window during which the average power should remain under the threshold, covering domains such as the CPU package and DRAM. It operates via low-level Model Specific Registers (MSRs), offering relatively low overhead and access to real-time energy measurements~\cite{unknown}.

Despite these advantages, RAPL remains a \textit{black-box} system with undocumented internal logic and varying behavior across platforms~\cite{amdmonitoring}. The actual power enforcement may suffer from saturation, overshoot, or platform-specific deviations~\cite{article2}. Moreover, traditional use of RAPL as static or rule-based modes fails to capture the dynamic behavior of HPC workloads. In practice, applications exhibit changing power profiles depending on whether the processor is performing computations, accessing memory, or handling data transfers. Because these patterns can vary across applications and over time, a fixed or rule-based power limit cannot adapt to such variations, often resulting in underutilized performance or energy inefficiency~\cite{petoumenos2015, amdmonitoring}.

Addressing these limitations requires adaptive strategies capable of responding to changing application and system conditions in real time. Autonomic computing provides a general framework for such adaptability, in which systems manage themselves according to high-level objectives through feedback loops~\cite{article4}. These loops may employ heuristic rules, artificial intelligence, machine learning, scheduling algorithms, or constraint programming. Within this paradigm, control theory offers a mathematically rigorous approach to regulating system behavior~\cite{article5,10766225,7194659}. Control based approaches have demonstrated effectiveness across a range of computing domains, including the management of cloud platforms~\cite{article6, berekmeri:hal-01297026}, energy and performance optimization in real-time systems~\cite{article7}, and coordinated operation in Internet of Things environments~\cite{IOT} or software~\cite{molina:hal-04885266}, \cite{zhao:hal-02509604}.

In HPC, the application of control theory to power regulation is relatively recent but has shown measurable benefits. Early contributions addressed workflow management, where feedback based control was applied to optimize the execution of jobs~\cite{10}, \cite{11}, \cite{13}%\BR{Find 1-2 references, other than ours to put here and replace \cite{12} as it is very old}
. More recent work has focused on energy regulation and power capping, where closed-loop controllers dynamically adjust system parameters based on real-time performance metrics~\cite{cerf2021}. %Several studies highlight the range of control strategies in HPC power management. %\KH{maybe separate the works that used RAPL and others that didnt to have a clear and well structured state of the art}
Only memory intensive workloads, \cite{cerf2021} implemented a PI controller and ~\cite{hawila2022} extended it with an adaptive algorithm. % and ~\cite{koudsThesis2023}  applied Model Predictive Control (MPC), 
Separately, only for compute intensive workloads, ~\cite{halitim:hal-05117585} proposed a cascaded robust control. While these approaches demonstrated promising results, they were tested only on very specific types of workloads.

Despite these advances, a clear limitation remains: most existing controllers are designed for only one workload type or phase,  which is highly unrealistic in practice where many HPC applications switch between different phases, each with its own power-performance characteristics. This creates a need for controllers that can adapt to changing workload phases while still meeting performance goals and staying within power limits. 

This paper focuses on extending the control framework \cite{cerf2021} %\BR{which control framework? ref?}
to handle varying workloads by proposing a unified controller capable of adapting across diverse execution phases particularly the transitions between compute and memory regimes. Our approach monitors the ratio between compute and memory activity in real time and adjusts the power cap accordingly. This enables smooth transitions between different workload phases and helps maintain an efficient trade-off between performance and energy consumption.

We propose and compare two control strategies that adapt the power cap in response to a workload indicator $\beta$, a parameter introduced to represent the ratio between compute and memory activity: %\KH{is this new parameter beta proposed by the author and and LPV identification is also one of the contributions ?}

\vspace{1mm}

\noindent
\textbf{• Gain scheduled PI control}: interpolates between two PI controllers tuned for the phases.

\vspace{1mm}

\noindent
\textbf{• Polytopic LPV $\mathcal{H}_\infty$ control}: robust controllers are synthesized at the operating extremes using Linear Matrix Inequalities (LMIs), then smoothly interpolated to handle intermediate workload conditions while ensuring stability and robustness against workload-dependent parameter variations.

Simulation results  demonstrate that the LPV controller provides improved tracking, robustness, and control smoothness under dynamic workload conditions, outperforming the PI baseline.
The remainder of this paper is organized as follows: Section~\ref{sec:system} introduces the system description. Section~\ref{sec:modeling} presents the models for memory and compute phases, identified from different benchmark applications. Section~\ref{sec:control} details the PI and LPV controllers. Section~\ref{sec:evaluation} reports simulation results and comparisons. Section~\ref{sec:conclusion} concludes and Section~\ref{sec:limitation} discusses limitation and future work.

\section{System Description} \label{sec:system}
\subsection{HPC Architecture and Workload Behavior}

\begin{figure}[H]
\centering
\includegraphics[width=0.35\textwidth]{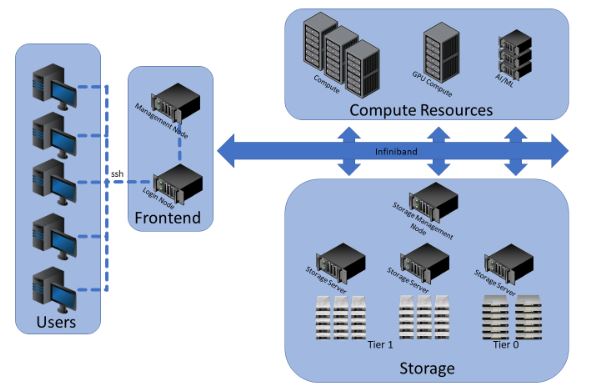}
\caption{\centering Typical HPC architecture}
\label{fig:hpc-architecture}
\end{figure}

%\KH{I think this could be merged into the introduction because i feel its more of a general HPC and workload behavior that already been mentioned in the intro}
HPC systems are composed of multiple interconnected layers that work together to execute complex, large-scale computations, as shown in Figure \ref{fig:hpc-architecture}. The job scheduler running on management nodes allocates tasks to the compute nodes, which may include general CPUs, GPUs, and specialized units. These compute resources are connected via a high-speed network, in parallel with storage resources. %~\cite{marquetteHPC}.

HPC workloads typically consist of multiple execution phases: %(Figure \ref{fig:workload-phases}).
\textbf{Compute-intensive} phases rely heavily on CPU or GPU resources to perform floating-point operations; \textbf{Memory-intensive} phases demand high bandwidth and low latency access to memory; \textbf{I/O-intensive} phases involve frequent data transfer to and from storage. Many applications exhibit a hybrid behavior, transitioning between different resource-intensive phases during execution \cite{2023} %\BR{ref?}
. These transitions significantly impact both power consumption and performance efficiency. Identifying and adapting to such workload variations in real time is essential for efficient dynamic power regulation. %\KH{i might be wrong but i think that having a figure for the previous paragraph (HPC architec, that shows the resources is more helpful than showing the phases in a blocks diagram)}

\subsection{Feedback Control Architecture} 

%\KH{Since this was first implemented in previous works, cite it and say that we use their control framework, so the reader know that this isnt the contribution of this work} 

\begin{figure}[!htbp]
    \centering
    
        \centering
        \includegraphics[width=0.82\linewidth]{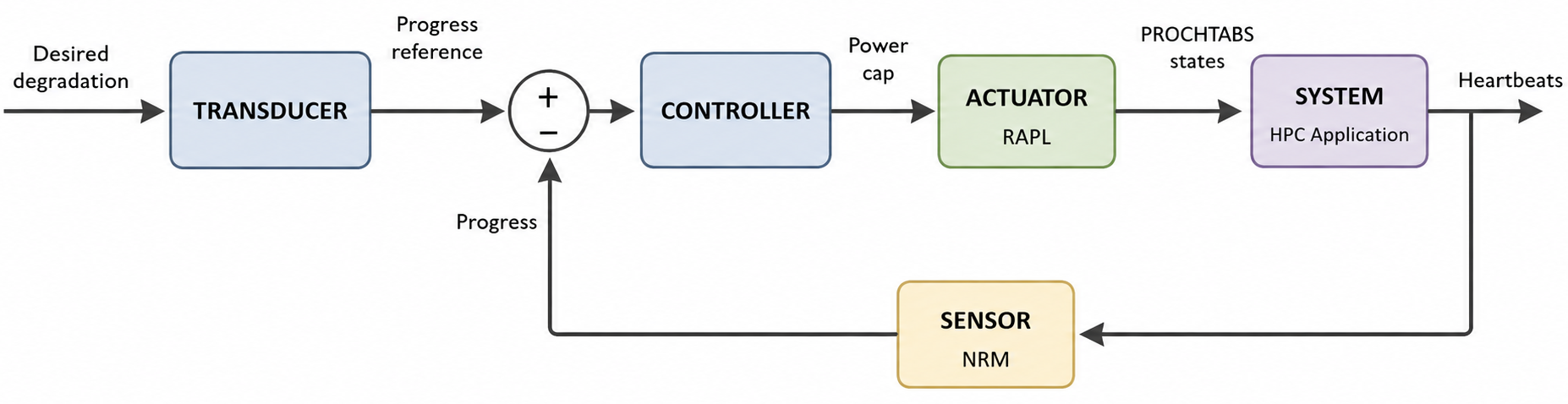}

    \caption{\centering Overall control architecture }
    \label{fig:overall-archi}
\end{figure}

The control architecture used in this study builds upon the experimental framework introduced in~\cite{cerf2021}. It implements a closed-loop feedback control mechanism that dynamically regulates processor power through runtime performance monitoring. A lightweight signal (known as the heartbeat) is generated periodically by the running application, representing the completion of a basic computational unit, such as a loop iteration or simulation step. These heartbeats are collected by the \textbf{Node Resource Manager (NRM)}, which functions as a middleware layer interfacing between the operating system, application, and control infrastructure (illustrated in Figure~\ref{fig:overall-archi}). The purpose of the heartbeat is to allow the system to monitor how quickly the application is making progress during execution.
 %\KH{I would prefer the term filter here but im not sure}
 To extract a reliable performance indicator from this raw data, the NRM includes a \textbf{progress sensor}. This sensor computes the application's execution rate by measuring the intervals between consecutive heartbeats. To reduce the effect of transient fluctuations caused by I/O activity and measurement noise, the measured progress signal is smoothed using a median-based filtering strategy \cite{cerf2021}.

\begin{comment}
    
\begin{equation}
    y(t_i) = \text{median} \left( \left\{ \frac{1}{t_k - t_{k-1}} \;\middle|\; t_k \in [t_{i-1}, t_i] \right\} \right)
\end{equation}
\end{comment}

%\BR{this figure 2 is quite difficult to follow, can we do a merge between the two figures? let's talk online !}

To balance performance and energy, a user-defined degradation tolerance specifies the maximum acceptable slowdown relative to the uncapped execution. The corresponding progress reference is computed as a fraction of the nominal rate (e.g., a 10 \% tolerance sets the target progress to 90 \% of the nominal value). The \textbf{controller} receives both the measured progress and the reference and computes the error. Based on this error, it generates a control signal representing a new power cap, which is applied to the processor using the \textbf{RAPL} interface. RAPL enforces the specified power cap by limiting the average power consumed over a fixed time window, effectively throttling the processor's energy use. Once the new cap is applied, the HPC application adapts to the available power, and its performance evolves accordingly. The resulting heartbeat rate is fed back into the control loop, allowing the system to continuously monitor and adjust power in response to changes in workload behavior. This feedback structure enables real-time control of energy consumption, ensuring that performance remains within defined bounds while optimizing power usage across diverse execution phases. %\KH{This section was a good place to refer to  2b}

\section{System Modeling} \label{sec:modeling}

\subsection{Workload Scheduling Parameter}

Dynamic power control in HPC systems relies on identifying the type of workload currently being executed. This requires classifying the application's execution phase in real time. Previous research~\cite{DynamicPowerOptimization2020} has demonstrated that combinations of hardware counters can reliably indicate execution phases. For example, high cache miss rates are commonly associated with memory bound workloads. Elevated instruction and branch activity may suggest mixed or hybrid phases. Meanwhile, low I/O sensor activity often corresponds to CPU focused execution or compute bound, whereas sudden bursts in I/O indicate disk or communication heavy behavior.

These indicators are often processed using dimensionality reduction and classification techniques, such as Principal Component Analysis (PCA), threshold-based methods, or decision-tree approaches, in order to identify transitions between workload phases during execution. Once the measured indicators stabilize over time, the system can assume that a new execution phase has been reached and adapt the control strategy accordingly.

To represent these workload variations mathematically, a dimensionless scheduling parameter $\beta \in [0,1]$ is introduced. $\beta=0$ represent a purely memory  workload, whereas $\beta=1$ represents a purely compute workload. Intermediate values describe mixed operating conditions between these two regimes. This parameter is used to interpolate the models and schedule the controller, allowing the control framework to dynamically adjust to workload shifts. In this study, $\beta$ is assumed to be available at each control update without delay. In practice, it can be estimated from hardware performance counters or provided by an external workload classifier \cite{DynamicPowerOptimization2020}. However, the design and validation of this estimator are beyond the scope of this work.

\subsection{System Identification }
\begin{figure}[!htbp]
    \centering
    
        \centering
        \includegraphics[width=0.62\linewidth]{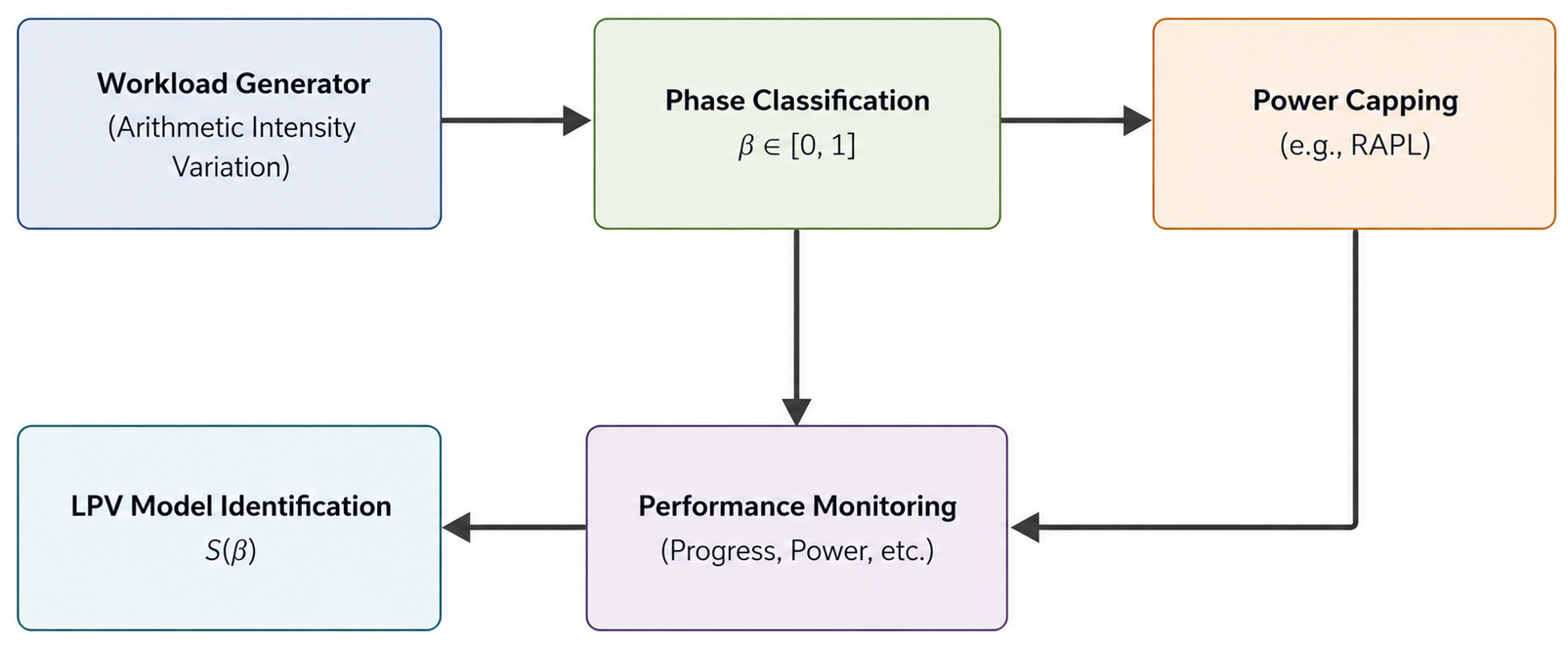}

    \caption{\centering LPV system identification methodology.}
  \label{fig:lpv-identification}
\end{figure}

The scheduling parameter $\beta$ enables the formulation of a Linear Parameter-Varying (LPV) model whose dynamics evolve with the workload phase. The objective is to identify a model $S(\beta)$ that relates the system progress $y(t)$, to the control input $u(t)$, corresponding to the applied processor power cap, under different workload conditions characterized by $\beta \in [0,1]$ (see Figure~\ref{fig:lpv-identification}).

% \KH{I think this sentense is not so clear on the variables (output: progress, workload variations: beta, control action: power caps)}

% this part to be modified later
This modeling process involves running benchmarks with controlled arithmetic intensity \footnote{Arithmetic intensity refers to the ratio of computational operations to memory accesses during a program's execution}%~\cite{eyraud2022symmetric}} 
profiles and injecting a range of power cap values. By varying this ratio, one can simulate different workload phases and observe how the system responds under power constraints. The resulting progress and power data is then used to identify models that vary smoothly with the scheduling parameter.%, forming the basis for a unified LPV control strategy. However, due to practical limitations such as the lack of programmable mixed-phase benchmarks and restricted access to hardware, we adopt a simplified modeling strategy. In this approach, compute and memory phases are treated as two separate systems \, control strategies are then interpolated based on the estimated workload phase. In future work, as more experimental data becomes available, we plan to revisit full LPV system identification and construct a globally valid model.

\mysubsubsection{Memory Intensive Workload Model}

For memory-bound workloads, experiments were conducted using the STREAM benchmark. Different levels of processor power caps were applied through the RAPL interface while recording the corresponding application progress using the heartbeat mechanism described previously. By varying the power cap input $P_{cap}$ and measuring the resulting progress $y$, a nonlinear relationship between power and performance was identified in ~\cite{hawila2022}:

\begin{equation}
\text{Progress} = K_L \left(1 - e^{-\alpha(a \cdot \text{Pcap} + b - \beta)} \right)
\label{eq:nonlinear-memory-model}
\end{equation}

\noindent where the parameters are summarized in Table~\ref{tab:memory-model-params}.

\begin{table}[h]
\centering
\caption{Model Parameters and Variables}
\label{tab:memory-model-params}

\footnotesize
\setlength{\tabcolsep}{4pt}
\renewcommand{\arraystretch}{1.0}

\begin{tabular}{|c|l|c|}
\hline
\textbf{Symbol} & \textbf{Description} & \textbf{Value} \\
\hline
$K_L$ & Maximum achievable progress & $42.4\,\mathrm{Hz}$ \\
$\alpha$ & Exponential decay-rate parameter & $0.032\,\mathrm{W}^{-1}$ \\
$a$ & RAPL actuator slope coefficient & $0.94$ \\
$b$ & RAPL actuator offset & $0.17\,\mathrm{W}$ \\
$\beta$ & Workload-specific offset parameter & $34.8\,\mathrm{W}$ \\
\hline
\end{tabular}

\end{table}

The nonlinear relation in~\eqref{eq:nonlinear-memory-model} characterizes the steady-state dependence between the applied power cap and the application progress. However, for controller synthesis, a dynamic approximation of the system behavior is required.

To obtain a control-oriented model, local identification experiments were performed around a high-power operating region ($P_{cap} \geq 70\,\mathrm{W}$), corresponding to the practical operating range typically used in performance-oriented HPC execution. In this region, the system behavior becomes smoother and can be reasonably approximated by a low-order linear model.

Using step variations of the power cap input and measuring the corresponding progress response, the memory-intensive workload dynamics were approximated by the following first-order transfer function:

\begin{equation}
G_m(s)=\frac{0.6269s+0.01798}{s+0.05317}
\label{eq:memory-transfer}
\end{equation}

\begin{figure}[!htbp]
    \centering
    \includegraphics[width=0.40\textwidth]{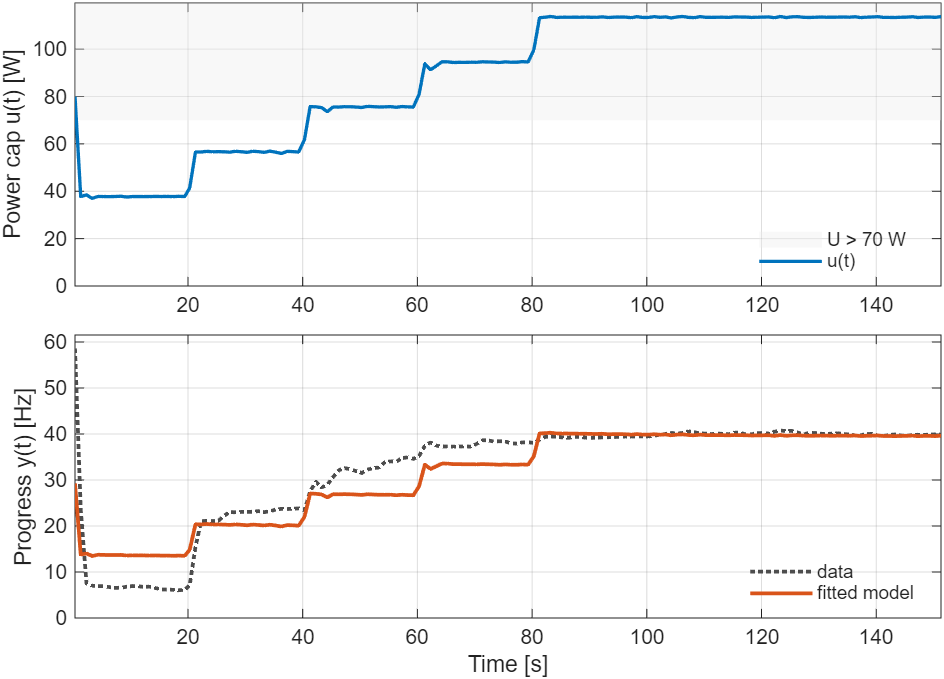} % adjust 0.45 as needed
        \vspace{-0.3cm}
    \caption{\centering Memory Model %\KH{I am not so sure i understand the pkg0 and 1 models + add on a parallel y the corresponding power signals}
    }
    \vspace{-0.2cm}
    \label{fig:memory-linearization}
\end{figure}

\noindent Figure~\ref{fig:memory-linearization} illustrates the measured experimental response together with the identified model approximation for the memory-intensive workload.

\mysubsubsection{Compute Intensive Workload Model}

\begin{figure}[!htbp]
    \centering
    \includegraphics[width=0.40\textwidth]{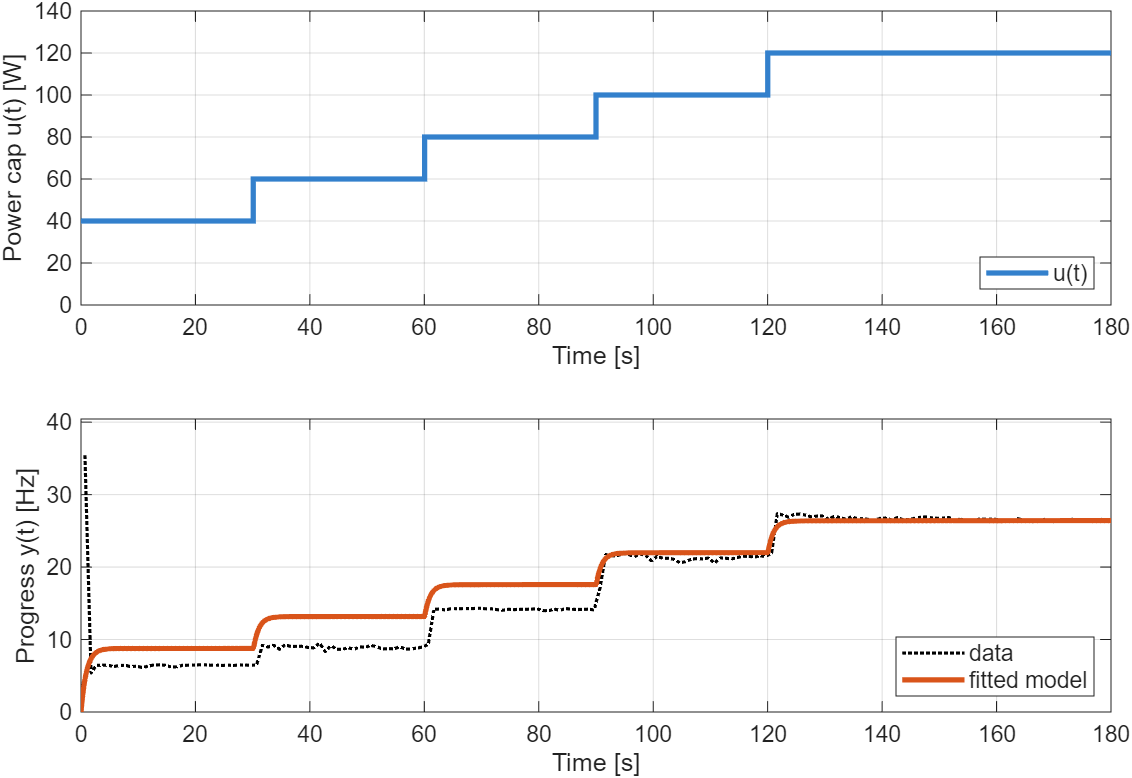} % adjust 0.45 as needed
            \vspace{-0.3cm}
    \caption{\centering Compute Model %\KH{I think it is better to use the dynamic model as u did for the memory, and i think the data is not the correct one of dahu}
    }
    \label{fig:compute}
\end{figure}

In the case of compute-intensive applications, the Embarrassingly Parallel (EP) benchmark is used. The relationship between the processor power cap and the application progress is illustrated in Figure~\ref{fig:compute} and discussed in~\cite{halitim:hal-05117585}. Although variations in the static gain can be observed across operating conditions, the system behavior is approximated here using a fixed low-order linear transfer function in order to obtain a tractable control-oriented model suitable for controller synthesis.

Using the same identification procedure as for the memory-intensive workload, the impact of the applied power cap on the application progress is approximated by the following transfer function:
\begin{equation}
G_c(s) = \frac{0.2361s + 0.4722}{s + 1.783}
\label{eq:compute-transfer-function}
\end{equation}

%here i can include another figure for compute 

\section{Control Design} \label{sec:control}

\subsection{Gain Scheduling PI}

The control objective is to track a target application progress while minimizing power consumption under varying workload conditions. The proposed controller employs a gain scheduling strategy based on the identified memory-intensive and compute-intensive models. The controller structure follows a classical Proportional-Integral (PI) control law:

\begin{equation}
u(t) = K_P \, e(t) + K_I \int_0^t e(\tau)\, d\tau
\label{eq:pi-control}
\end{equation}

\noindent where $e(t)=r(t)-y(t)$ denotes the tracking error between the reference progress $r(t)$ and the measured progress $y(t)$.

Each PI controller is tuned separately to satisfy standard performance requirements ensuring both responsiveness and closed-loop stability. The adopted specifications are summarized in Table~\ref{tab:pi_specs_gains}. The tuning objectives aim to ensure fast reference tracking while avoiding excessive power fluctuations that could destabilize the computing platform.

\begin{table}[H]
  \centering
  \scriptsize
  \caption{PI controller specifications and gains}
  \label{tab:pi_specs_gains}
  \footnotesize
  \renewcommand{\arraystretch}{1.1}
  \begin{tabular}{|l|c|}
    \hline
    \textbf{Metric / Parameter} & \textbf{Value} \\
    \hline
    \multicolumn{2}{|c|}{\textit{Performance specifications}} \\
    \hline
    Rise time         & $\leq 1.5\,\mathrm{s}$ \\
    Settling time     & $\leq 3.0\,\mathrm{s}$ \\
    Maximum overshoot & $\leq 10\%$ \\
    \hline
    \multicolumn{2}{|c|}{\textit{Tuned PI gains}} \\
    \hline
    Memory intensive $(K_P, K_I)$  & $(0.0213,\;0.0397)$ \\
    Compute intensive $(K_P, K_I)$ & $(4.3161,\;8.1339)$ \\
    \hline
  \end{tabular}
\end{table}

The gain scheduling mechanism employs linear interpolation of the controller gains according to the scheduling parameter $\beta$. The scheduled gains are computed as

\begin{equation}
\left\{
\begin{aligned}
K_P(\beta) &= (1 - \beta) K_{P,\text{mem}} + \beta K_{P,\text{comp}}, \\
K_I(\beta) &= (1 - \beta) K_{I,\text{mem}} + \beta K_{I,\text{comp}}.
\end{aligned}
\right.
\label{eq:gain-scheduling}
\end{equation}

\subsection{LPV Control Design}

The polytopic LPV approach represents the system as a convex interpolation between the memory-intensive and compute-intensive operating conditions using the scheduling parameter $\beta \in [0,1]$. The transfer functions given in~\eqref{eq:memory-transfer} and~\eqref{eq:compute-transfer-function} are first converted into state-space representations. The resulting LPV model is written as

\begin{equation*}
\scriptsize
\begin{pmatrix}
A(\beta) & B(\beta) \\
C(\beta) & D(\beta)
\end{pmatrix}
=
\sum_{i=1}^{N} \alpha_i(\beta)
\begin{pmatrix}
A_i & B_i \\
C_i & D_i
\end{pmatrix},
\qquad
\sum_{i=1}^{N} \alpha_i(\beta)=1
\label{eq:lpv-boxed}
\end{equation*}

\noindent where $N=2$ corresponds to the two operating regimes and the interpolation coefficients are defined as

\[
\alpha_1(\beta)=1-\beta,
\qquad
\alpha_2(\beta)=\beta.
\]

The objective is to design an LPV controller $K(\beta)$ such that the closed-loop transfer from the exogenous input $w$ to the regulated output $z$ satisfies

\[
\|T_{zw}\|_{\infty} < \gamma_{\infty},
\]

\noindent ensuring bounded disturbance amplification and closed-loop robustness.

The $\mathcal{H}_{\infty}$ formulation incorporates two weighting functions: $W_e(s)$ to penalize tracking errors and $W_u(s)$ to limit excessive control activity. The regulated output is defined as

\[
z =
\begin{bmatrix}
W_e e \\
W_u u
\end{bmatrix},
\]

\noindent where $e(t)=r(t)-y(t)$ is the tracking error and $u(t)$ is the applied power cap.

For each vertex $i \in \{1,2\}$ corresponding to the memory-intensive and compute-intensive operating conditions, the augmented plant $P_i$ is represented as

\begin{equation}
\begin{bmatrix}
\dot{x}_P \\ z \\ y
\end{bmatrix}
=
\begin{bmatrix}
A_{P,i} & B_{w,i} & B_{u,i} \\
C_{z,i} & D_{zw,i} & D_{zu,i} \\
C_{y,i} & D_{yw,i} & D_{yu,i}
\end{bmatrix}
\begin{bmatrix}
x_P \\ w \\ u
\end{bmatrix},
\label{eq:augmented_plant}
\end{equation}

where $x_P$ includes both the plant states and the weighting-filter dynamics.

\begin{figure}[!htbp]
    \centering
    \includegraphics[width=0.38\textwidth]{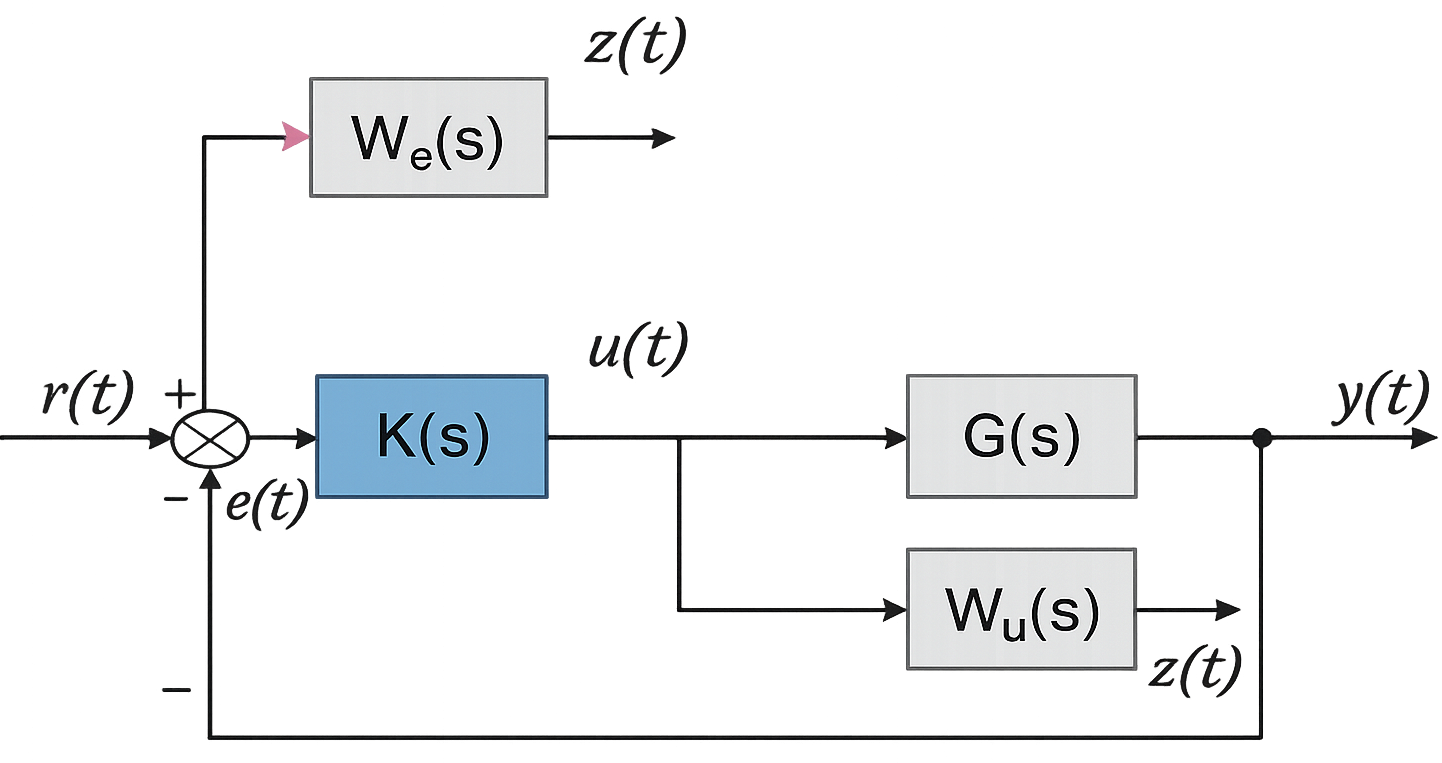}
    \caption{Standard $\mathcal{H}_{\infty}$ control structure used for LPV controller synthesis.}
    \label{fig:h}
\end{figure}

\vspace{3pt}
\noindent
Using the standard bounded-real formulation for polytopic LPV systems~\cite{15,16}, the controller synthesis problem can be expressed as a set of Linear Matrix Inequalities (LMIs) evaluated at each polytope vertex.

The existence of a controller $K(\beta)$ satisfying the prescribed $\mathcal{H}_{\infty}$ performance level is guaranteed if there exist symmetric positive definite matrices $X>0$ and $Y>0$, together with auxiliary matrices $A_t^i$, $B_t^i$, $C_t^i$, and $D_t^i$ representing transformed decision variables used to linearize the synthesis problem, such that for each vertex $i$ the following LMI holds:

\begin{equation}
\mathbf{M}_i =
\begin{bmatrix}
M_{11}^i & (M_{21}^i)^{\!\top} & (M_{31}^i)^{\!\top} & (M_{41}^i)^{\!\top} \\[3pt]
M_{21}^i & M_{22}^i & (M_{32}^i)^{\!\top} & (M_{42}^i)^{\!\top} \\[3pt]
M_{31}^i & M_{32}^i & -\gamma_{\infty}I & (M_{43}^i)^{\!\top} \\[3pt]
M_{41}^i & M_{42}^i & M_{43}^i & -\gamma_{\infty}I
\end{bmatrix} < 0,
\label{eq:lmi}
\end{equation}
with
\begin{align*}
M_{11}^i &= A_{P,i}X + X A_{P,i}^{\!\top} + B_{u,i}C_t^i + (B_{u,i}C_t^i)^{\!\top},\\
M_{21}^i &= A_t^i + (A_{P,i} + B_{u,i}D_t^iC_{y,i})^{\!\top},\\
M_{22}^i &= Y A_{P,i} + A_{P,i}^{\!\top} Y + B_t^i C_{y,i} + (B_t^i C_{y,i})^{\!\top},\\
M_{31}^i &= (B_{w,i} + B_{u,i}D_t^iD_{yw,i})^{\!\top},\\
M_{32}^i &= (Y B_{w,i} + B_t^iD_{yw,i})^{\!\top},\\
M_{41}^i &= C_{z,i}X + D_{zu,i}C_t^i,\\
M_{42}^i &= C_{z,i} + D_{zu,i}D_t^iC_{y,i},\\
M_{43}^i &= D_{zw,i} + D_{zu,i}D_t^iD_{yw,i}.
\end{align*}

\noindent The resulting LMIs are convex and can be efficiently solved using semidefinite programming solvers such as SeDuMi or SDPT3.

The controller matrices are then reconstructed from the auxiliary variables using the standard back-transformation,

\[
K_i =
\begin{bmatrix}
A_c^i & B_c^i \\
C_c^i & D_c^i
\end{bmatrix},
\]

and the global controller is obtained by interpolation:

\begin{equation}
K(\beta)=\sum_{i=1}^{N}\alpha_i(\beta)K_i
\label{eq:controller-interpolation}
\end{equation}

\section{Evaluation} \label{sec:evaluation}

\subsection{PI Controller evaluation} %\label{}

The gain scheduled PI controller was implemented and evaluated under different workload conditions and reference progress levels representative of HPC power-capping scenarios on the Dahu cluster environment~\cite{cerf2021}. 

\begin{figure}[!htbp]
    \centering
    \includegraphics[width=\columnwidth]{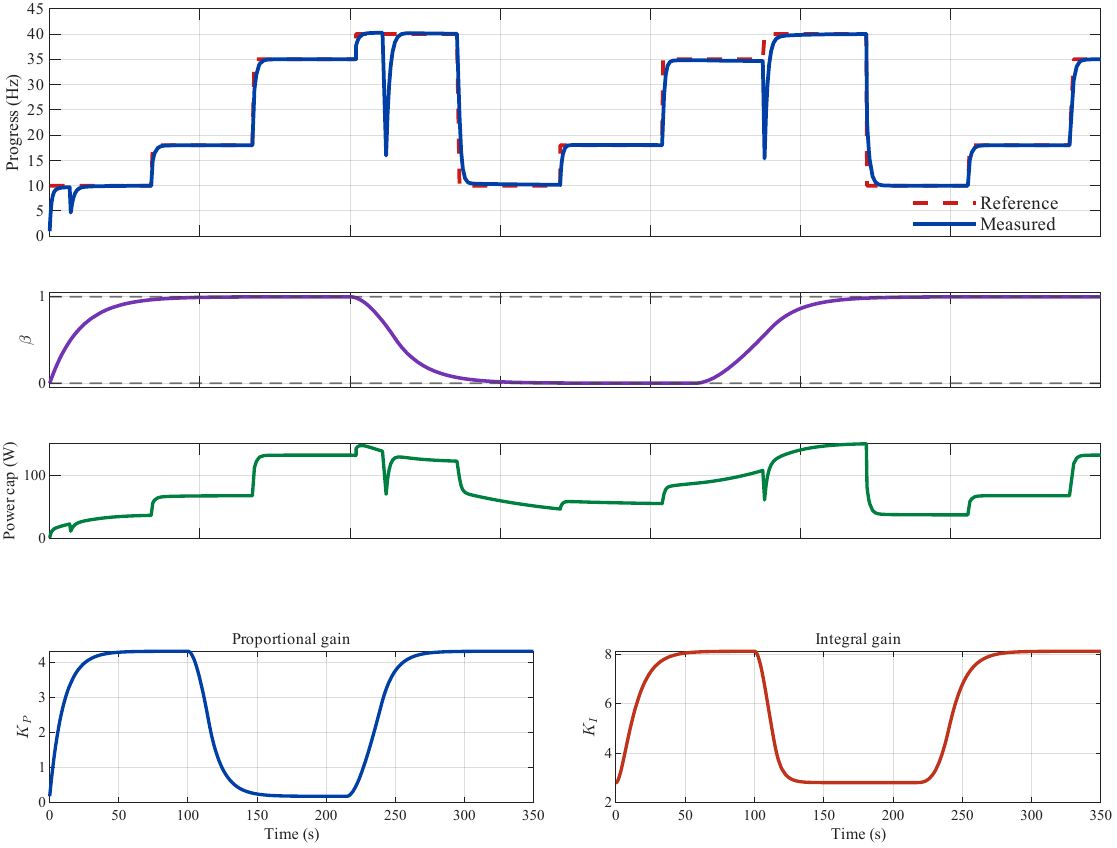}
    \caption{Closed-loop response of the gain scheduled PI controller under smooth workload variations.}
    \label{fig:pi_smooth}
\end{figure}

Figure~\ref{fig:pi_smooth} illustrates the closed-loop behavior of the gain scheduled PI controller. 
The first subplot shows the progress tracking, where the measured progress $y(t)$ closely follows the reference signal.
The second subplot presents the scheduling parameter $\beta$. 
The third subplot displays the control input $u(t)$, corresponding to the applied processor power cap (in watts). The fourth subplot presents the tracking error, while the last two subplots show the proportional and integral gains, $K_p(\beta)$ and $K_i(\beta)$ variations, respectively.

Overall, the figure demonstrates that the controller effectively maintains the desired progress while dynamically adjusting its gains and power cap according to workload variations. The tracking error remains bounded and rapidly converges toward zero after each workload transition.

\begin{comment}

\begin{figure}[!htbp]
    \centering
    \includegraphics [width=0.5\textwidth]{PInrml (1).jpg}
    \caption{Gain scheduling PI performance under smooth workload variations.}
    \label{fig:pi_performance}
\end{figure}

\begin{figure}[!htbp]
    \centering
    \includegraphics[width=0.45\textwidth]{PIERROR (2).jpg}
    \caption{Tracking error analysis of the gain-scheduling PI controller. 
The plots show the time evolution and statistical distribution of the tracking error}

    \label{fig:pi_error}
\end{figure}
\end{comment}

Despite these positive characteristics, the main limitation of the gain scheduled PI controller lies in the occurrence of abrupt gain variations when the workload parameter $\beta$ changes rapidly. This effect is particularly visible during transitions between operating conditions, where sharp gain variations induce transient tracking deviations and short periods of instability.

\begin{figure}[!http]
    \centering
    \includegraphics[width=\columnwidth]{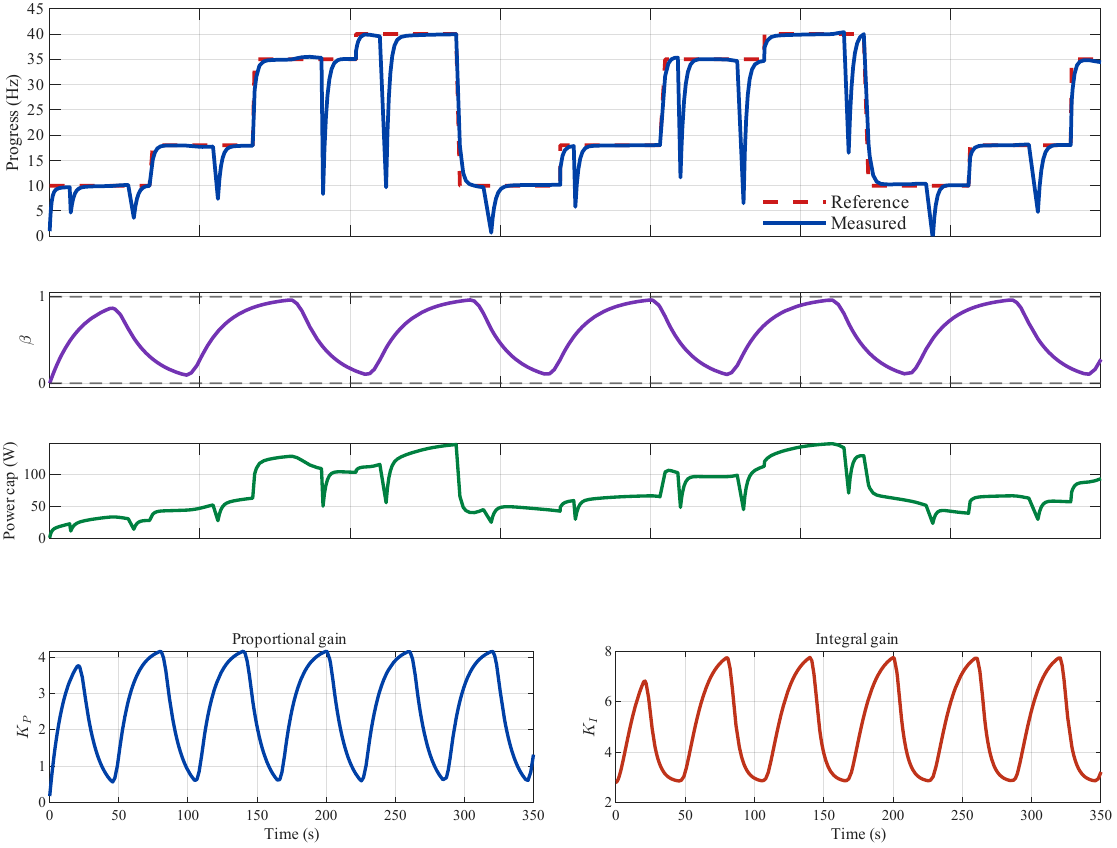}
    \caption{Closed-loop response of the gain scheduled PI controller under aggressive workload transitions.}
    \label{fig:pi_aggressive}
\end{figure}

To further analyze this behavior, the same experiment was repeated with sharper transitions in $\beta$, reproducing frequent and abrupt workload changes. Under these conditions, the tracking accuracy decreases noticeably and larger error spikes emerge, as illustrated in Figure~\ref{fig:pi_aggressive}. This highlights the sensitivity of gain scheduled PI control to rapid variations in $\beta$ and motivates the use of an LPV formulation to ensure smoother and more robust closed-loop performance under dynamic workload conditions.

\subsection{LPV Controller evaluation}

\begin{figure}[!http]
    \centering
    \includegraphics[width=\columnwidth]{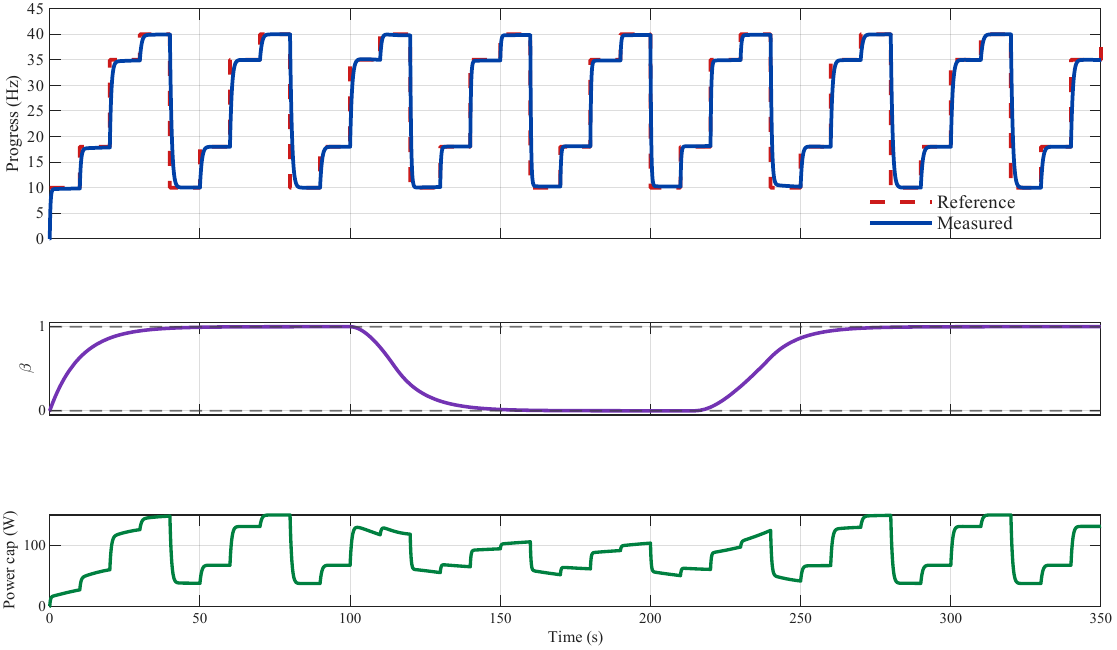}
    \caption{Closed-loop response of the LPV $\mathcal{H}_{\infty}$ controller under smooth workload variations.}
    \label{fig:lpv_smooth}
\end{figure}

The effectiveness of the proposed LPV controller is validated through simulation results. Figure~\ref{fig:lpv_smooth} illustrates the closed-loop behavior of the LPV $\mathcal{H}_{\infty}$ controller under smooth workload variations. The controller successfully tracks the desired progress reference while smoothly adapting to workload variations represented by the scheduling parameter $\beta$. Compared with the gain scheduled PI controller shown in Figure~\ref{fig:pi_smooth}, the LPV controller exhibits smaller tracking-error and smoother transient behavior, confirming the effectiveness of the polytopic LPV formulation.

\begin{figure}[!http]
    \centering
    \includegraphics[width=\columnwidth]{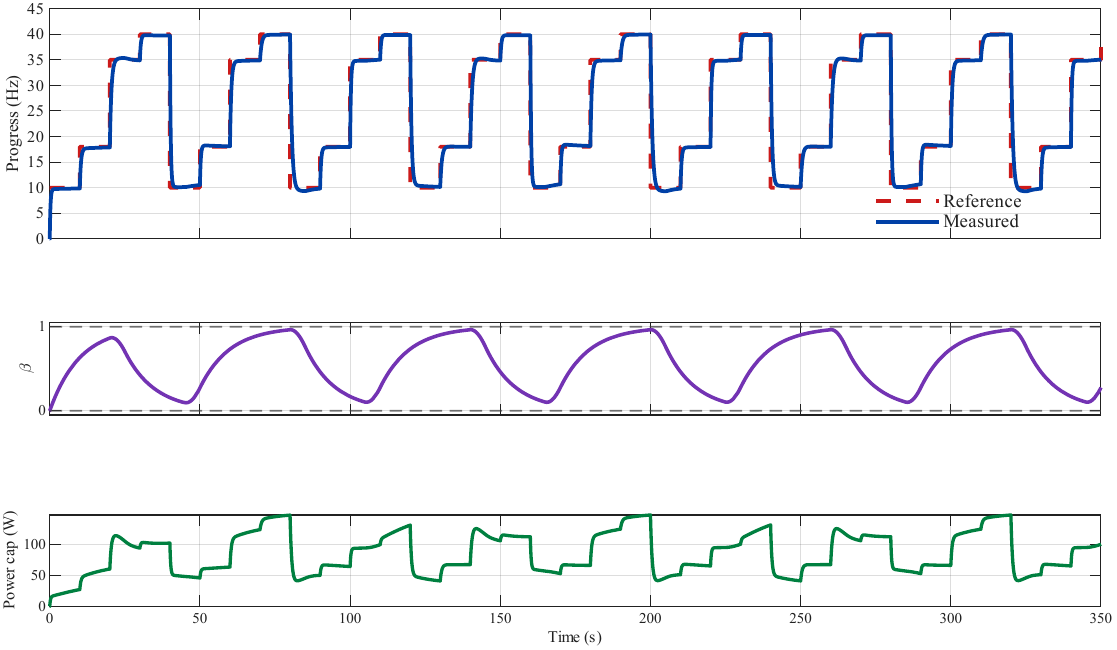}
    \caption{Closed-loop response of the LPV $\mathcal{H}_{\infty}$ controller under aggressive workload transitions.}
    \label{fig:lpv_aggressive}
\end{figure}

Concerning aggressive workload transitions, the results are presented in Figure~\ref{fig:lpv_aggressive}. Compared with the gain scheduled PI controller shown in Figure~\ref{fig:pi_aggressive}, the LPV controller maintains smoother closed-loop behavior and improved reference tracking despite rapid variations in the scheduling parameter $\beta$. The reduction of tracking-error peaks is clearly visible in the error subplot, highlighting the improved robustness of the LPV formulation under highly dynamic workload conditions.

%To quantify tracking performance, Figure~\ref{fig:lpverror} provides a detailed error analysis. The left plot illustrates the evolution of tracking error over time which indicate  that the error remains well controlled  even under aggressive transitions in reference progress and highly dynamic changes in workload type as in Figures~\ref{fig:lpvaggressive} and~\ref{fig:lpvaggrerror} 

\begin{comment}
\begin{figure}[!htbp]
\centering
\includegraphics[width=0.95\linewidth]{agrilpv.jpg}
\caption{LPV Controller Performance under aggressive dynamics}
\label{fig:lpvaggressive}
\end{figure}

\begin{figure}[!htbp]
\centering
\includegraphics[width=0.9\linewidth]{lpvaggrerror.jpg}
\caption{Tracking Error Analysis under aggressive conditions.}
\label{fig:lpvaggrerror}
\end{figure}

\end{comment}

\begin{figure}[!htbp]
    \centering
    \includegraphics[width=0.45\textwidth]{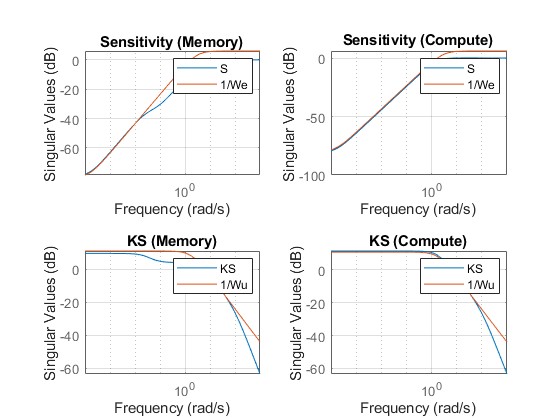}
    \caption{\centering Sensitivity function for the LPV \( \mathcal{H}_\infty \) controller.}
    \label{fig:sensitivity}
\end{figure}

We see from the previous figures that the LPV \( \mathcal{H}_\infty \) controller performs well under both smooth and aggressive workload variations. To further validate the quality of the controller, we also analyzed its frequency-domain behavior using sensitivity functions. The plots of the sensitivity \( S \) and complementary sensitivity \( KS \) functions (Figure~\ref{fig:sensitivity}) show that the controller responses stay well within the designed template. Specifically, \( S \) remains below the inverse of the weighting function \( 1/W_e \), indicating good disturbance rejection in the low-frequency range. Similarly, \( KS \) remains below \( 1/W_u \), showing that the control effort is bounded and does not overreact at high frequencies. %These results confirm that the \( \mathcal{H}_\infty \) performance level \( \ga

%\subsection{Comparison %\BR{with other simple PI and adaptive Pi and th eMPC of Kouds if possible} i dont  think  it is possible  because  the  MPC and adaptive are only for  memory  case system }

\begin{table}[H]
  \centering
  \caption{Performance comparison of PI and LPV $\mathcal{H}_\infty$ controllers.}
  \label{tab:controller_comparison}
  \footnotesize
  \renewcommand{\arraystretch}{1.2}
  \begin{tabular}{|l|cc|cc|}
    \hline
    \multirow{2}{*}{\textbf{Metric}} & 
    \multicolumn{2}{c|}{\textbf{Simple Workload}} & 
    \multicolumn{2}{c|}{\textbf{Aggressive Workload}} \\ 
    \cline{2-5}
    & \textbf{PI} & \textbf{LPV $\mathcal{H}_\infty$} & \textbf{PI} & \textbf{LPV $\mathcal{H}_\infty$} \\
    \hline
    Fit (\%) & 89.9 & \textbf{92.9} & 78.7 & \textbf{92.7} \\
    RMSE & 2.69 & \textbf{1.63} & 5.32 & \textbf{1.76} \\
    \hline
  \end{tabular}
\end{table}

\begin{comment}

\begin{table}[H]
  \centering
  \caption{Performance comparison of PI and LPV $\mathcal{H}_\infty$ controllers}
  \label{tab:controller_comparison}
  \footnotesize
  \renewcommand{\arraystretch}{1.1}
  \begin{tabular}{|l|c|c|}
    \hline
    \textbf{Metric} & \textbf{PI} & \textbf{LPV $\mathcal{H}_\infty$} \\
    \hline
    \multicolumn{3}{|c|}{\textit{Simple workload}} \\
    \hline
    Fit (\%) & 89.9 & 92.9 \\
    RMSE & 2.69 & 1.63 \\
    \hline
    \multicolumn{3}{|c|}{\textit{Aggressive workload}} \\
    \hline
    Fit (\%) & 78.7 & 92.7 \\
    RMSE & 5.32 & 1.76 \\
    \hline
  \end{tabular}
\end{table}

\end{comment}

A synthetic comparison is given in Table~\ref{tab:controller_comparison} which demonstrates that while both controllers perform adequately under simple conditions, the LPV controller maintains significantly better performance and superior robustness to dynamic operating conditions, during aggressive workload variations, similar to those of real-world applications. %, with substantially lower tracking error .

\section{Conclusion} \label{sec:conclusion}

This paper presented a dynamic power-capping framework for mixed HPC workloads using two feedback strategies: a gain scheduled PI controller and a polytopic LPV $\mathcal{H}_\infty$ controller scheduled by a workload indicator $\beta $ that interpolates between memory and compute bound phases. The models for both regimes were identified from benchmark data and evaluated in simulation, considering realistic constraints on the power cap input $u$ [Watts], and on the progress output $y$ [Hz]. 

Across tracking and phase switching scenarios, both controllers achieved effective regulation and respected power limits. However, the gain scheduled PI controller lacks formal stability guarantees when $\beta$ varies rapidly, leading to transient oscillations and unpredictable performance during abrupt workload transitions. In contrast, the LPV $\mathcal{H}_\infty$ controller, designed via convex LMI synthesis, ensures stability and smooth behavior across the entire operating range. It consistently achieved lower tracking error, reduced control variance, and smoother transients, demonstrating superior robustness to fast workload changes. 

These results confirm that polytopic LPV  control provides a reliable choice for dynamic power regulation in HPC environments with highly variable workload phases.

\section{Limitations \& Future work} \label{sec:limitation}
This work assumes a provided scheduling parameter $\beta$ and relies on simulation based evaluation using simplified SISO models. While this approach allows for some analysis and controller comparison, it does not fully capture the nonlinear dynamics, and potential delays of the actual RAPL actuation on real hardware. 
Actual developments are focusing on deriving $\beta$ online from hardware performance counters and extending the LPV identification framework to cover broader operating regions and include delay effects. 
Additional work will aim to validate the proposed control strategies on physical HPC nodes and multi-node setups, integrating real-time power and performance measurements. 
Finally, we plan to investigate constraint-aware and distributed LPV control formulations to ensure safe and coordinated power regulation across multiple heterogenous computing units.

 \label{sec:future}

\begin{comment}

\begin{ack}
Place acknowledgments here.
\end{ack}
\end{comment}
%\section*{DECLARATION OF GENERATIVE AI AND AI-ASSISTED TECHNOLOGIES IN THE WRITING PROCESS}
%During the preparation of this work, the author used \textit{ChatGPT (OpenAI, GPT-5 model)} to assist with English editing, figure labeling, and LaTeX formatting.  
%After using this tool, the author carefully reviewed and edited the content to ensure accuracy and originality and takes full responsibility for the final version of the manuscript.

\bibliographystyle{IEEEtran}
\bibliography{ifacconf}
         % bib file to produce the bibliography
                                                     % with bibtex (preferred)
                                                   
%\begin{thebibliography}{xx}  % you can also add the bibliography by hand

%\bibitem[Able(1956)]{Abl:56}
%B.C. Able.
%\newblock Nucleic acid content of microscope.
%\newblock \emph{Nature}, 135:\penalty0 7--9, 1956.

%\bibitem[Able et~al.(1954)Able, Tagg, and Rush]{AbTaRu:54}
%B.C. Able, R.A. Tagg, and M.~Rush.
%\newblock Enzyme-catalyzed cellular transanimations.
%\newblock In A.F. Round, editor, \emph{Advances in Enzymology}, volume~2, pages
%  125--247. Academic Press, New York, 3rd edition, 1954.

%\bibitem[Keohane(1958)]{Keo:58}
%R.~Keohane.
%\newblock \emph{Power and Interdependence: World Politics in Transitions}.
%\newblock Little, Brown \& Co., Boston, 1958.

%\bibitem[Powers(1985)]{Pow:85}
%T.~Powers.
%\newblock Is there a way out?
%\newblock \emph{Harpers}, pages 35--47, June 1985.

%\bibitem[Soukhanov(1992)]{Heritage:92}
%A.~H. Soukhanov, editor.
%\newblock \emph{{The American Heritage. Dictionary of the American Language}}.
%\newblock Houghton Mifflin Company, 1992.

%\end{thebibliography}

\end{document}